\documentclass[prl,twocolumn,superscriptaddress]{revtex4-1}
\usepackage{amssymb,amsmath}
\usepackage{graphicx}
\usepackage[colorlinks=true,linkcolor=blue,citecolor=blue,urlcolor=blue]{hyperref}

\begin{document}

\title{Generalized Independent Component Analysis for Extracting Eigen-Modes of a Quantum System}

\author{Yadong Wu}
\affiliation{Institute for Advanced Study, Tsinghua University, Beijing, 100084, China}

\author{Hui Zhai}
\affiliation{Institute for Advanced Study, Tsinghua University, Beijing, 100084, China}

\begin{abstract}

In many dynamical probes of a quantum system, quite often multiple eigenmodes are excited. Therefore, the experimental data can be quite messy due to the mixing of different modes, as well as the background noise, despite that each mode manifests itself as a single frequency oscillation. Here we develop an unsupervised machine learning algorithm to extract the frequencies of these oscillations from such measurement data, that is, the eigenenergies of these modes. This method is particularly useful when the measurement time is not long enough to perform the Fourier transformation. Our method is inspired by the independent component analysis method and its application to the ``cocktail party problem". In that problem, the goal is to recover each voice from detectors that detect signals of many mixed voices, and the principle is to find out signals that possess features and are away from a Gaussian distribution. Instead, our generalization is to find out signals that are close to a single frequency oscillation. We demonstrate the advantage of our method by an example of analyzing the collective mode of cold atoms. We believe this method can find broad applications in analyzing data from dynamical experiments in quantum systems.      
\end{abstract}

\maketitle

In quantum systems, eigen-modes have discrete eigen-energies, and each eigen-mode with a fixed eigen-energy manifests itself as a single-frequency oscillation in a dynamical probe. The frequency of the oscillation, up to an $\hbar$, is the eigen-energy of the eigen-mode. Therefore it is quite common task in many quantum physics experiments to extract a single-frequency oscillation from dynamical measurements. If the measurement can be performed for sufficient long time, the frequency can be determined by the Fourier transformation. However, in many cases it is not possible to perform a long time measurement, either because the oscillation quickly damps out, or because the lifetime of the system is limited. An alternative way commonly used in practice is to fit the dynamics with a damped harmonic oscillator. Nevertheless, it is quite often that the probe excites multiply modes with different frequencies, and some times the oscillation is also embedded in noisy signals, therefore the fitting becomes not so reliable. Thus, when the measurement time is short, and the data contains multiple frequencies and is noisy, the task of extracting single-frequency oscillation becomes quite challenging. The purpose of this letter is to design a machine learning algorithm to solve this problem. 

To start with, let us first briefly review the independent component analysis (ICA) \cite{ICA}and its application to the classical ``cocktail party problem"\cite{ccp}. The ``cocktail party problem" considers $N$-number of sources of voices, each of which is a sequence of data denoted by $s_i(t), (i=1,\dots,N)$, and $M$-number of detectors with $M>N$, each of which obtains a signal $x_j(t), (j=1,\dots,M)$ as a linear combination of all $s_i(t)$. That is to say, there is a $M\times N$ matrix $A$ such that
\begin{equation}
x_j(t)=\sum\limits_{i}A_{ji}s_i(t). 
\end{equation}
The matrix $A$ is unknown and is assumed to be independent of $t$, and the goal is to find out $A^{-1}$ such that one can recover each voice $s_i(t)$ from the signals obtained by the detectors $\{x_j(t), (j=1,\dots,M)\}$. The aforementioned problem in a quantum system is actually the same as the ``cocktail party problem". Each single frequency oscillation from a given eigen-mode can be viewed as a source $s_i(t)$, and several different measurements $x_j(t)$ obtain several different superpositions of $s_i(t)$. To extract the frequency or eigen-energy of each mode, one first needs to recover $s_i(t)$ from the measurements $\{x_j(t)\}$. 

What lies behind the ICA application of this problem is the central limit theorem\cite{clt}. For each signal, when we perform statistics over certain duration of $t$, we can obtain the distribution for each signal, which always possesses certain feature. The central limit theorem says that when one adds up many such signals together, the distribution of $x_j(t)$ will approach a Gaussian distribution and looks like a noise. The signal with Gaussian distribution displays maximum entropy for given mean and variance. Thus, the working principle of ICA is to find out $A^{-1}$ such that the distribution of $s_i(t)=(A^{-1})_{ij}x_j(t)$ deviates from a Gaussian distribution as far as possible, or in a more quantitative description, that the entropy of $s_i(t)=\sum_j(A^{-1})_{ij}x_j(t)$ is as small as possible. 

To be more precise, we first normalize the data $x_j(t)$ such that its mean value $\bar{x}_j(t)=0$ and its covariance matrix $X^\dag X=I$, where the statistics is performed over a sufficiently long duration of $t$. We further require $A^\dag A=I$, which ensures the mean value of $s_i(t)$ is also zero and its variance also equals one. By performing statistics over $t$, for a given sequence $s(t)$, one can obtain its distribution $\mathcal{P}[s, u]$, where $u$ is the range of $s(t)$. Then, we can compute the entropy of a given sequence $s(t)$ as
\begin{equation}
\mathcal{H}[s]=-\int \mathcal{P}[s,u]\log\mathcal{P}[s,u] du \label{entropy}.
\end{equation}
If $s_\text{Gau}$ is a sequence obeying a Gaussian distribution with zero mean and unity variance, $H[s_\text{Gau}]=\frac{1}{2}-\frac{1}{2}\log\frac{1}{2\pi}$ is the entropy maximum for sequences with same mean and variance. Hence, the ICA method is to find out $A^{-1}$ such that for each $s_i(t)=\sum_j(A^{-1})_{ij}x_j(t)$,
\begin{equation}
J[s_i]=\mathcal{H}[s_\text{Gau}]-\mathcal{H}[s_i]
\end{equation}  
is maximized. In practices, since it is hard to directly compute $\mathcal{P}[s,u]$, several formula have been proposed to approximate $\mathcal{H}[s]$\cite{Happ}.

Below we will apply the ICA method to a dynamical probe which excites three different modes, mixed together with background noise. We will see that the ICA method can work but the outcome is not ideal. The reason that it does not work well can also be understood. Because the ICA only assumes that the signal has certain feature but does not full explore what exactly the feature is. This is good for original ``cocktail party problem" because it does not require prior knowledge of each voice. However, as discussed above, in our quantum problem each eigen-mode has a fixed energy and manifests as a single frequency oscillation, but this feature of being single frequency oscillation is not utilized in the ICA method above. Therefore, the main result of this work is to present a generalized ICA method that aims at finding out signal of single frequency oscillation, short-noted as \textit{s-ICA}. 

\textit{s-ICA.} In short, let us consider a reference signal $s_\text{ref}=\sqrt{2}\cos(\omega t)$. Here $\sqrt{2}$ is chosen to ensure that the variance equals one. The ICA method is to find out an $A^{-1}$ such that each $s_i=\sum_j(A^{-1})_{ij}x_j$ is away from $s_\text{Gau}$ as much as possible; and our s-ICA method is to find out an $A^{-1}$ such that each $s_i=\sum_j(A^{-1})_{ij}x_j$ is close to $s_\text{ref}$ as much as possible.  

To quantify how $s_i$ is close to $s_\text{ref}$, instead of using entropy we consider a quantity called the cumulants\cite{cum}. The cumulants is defined as
\begin{equation}
\mathcal{K}[s,z]=\log(\langle e^{z s(t)}\rangle),
\end{equation}
where $\langle \dots\rangle$ means performing average over a certain duration of $t$. 
For $s_\text{ref}$, it is straightforward to compute 
\begin{equation}
\mathcal{P}[s_\text{ref},u]=\frac{1}{\pi}\frac{1}{\sqrt{2-u^2}}, \label{Pref}
\end{equation}
for $-\sqrt{2}<u<\sqrt{2}$, and otherwise $\mathcal{P}[s_\text{ref},u]=0$, and then 
\begin{equation}
\langle e^{z s_\text{ref}(t)}\rangle=\int \mathcal{P}[s_\text{ref},u]e^{izu}du=I_0(\sqrt{2}z),
\end{equation}
and 
\begin{equation}
\mathcal{K}[s_\text{ref},u]=\log(I_0(\sqrt{2}z)). \label{Kref}
\end{equation}
It is very important to acknowledge that $\mathcal{P}[s_\text{ref},u]$, and consequently, $\mathcal{K}[s_\text{ref},u]$, is independent of the frequency $\omega$. Thus we can use $\mathcal{K}[s_\text{ref},u]$ as a reference to quantify how close $s_i$ is to $s_\text{ref}$ without knowing the value of $\omega$ as a prior. In principle, we should require $\mathcal{K}[s,z]$ to be close to $\mathcal{K}[s_\text{ref},z]$ for all $z$. In practices, we consider a set of $\{z_i\},i=1,\dots, L$, and define the loss function as 
\begin{equation}
\mathcal{L}[s]=\sum\limits_{i=1}^{L}(\mathcal{K}[s,z_i]-\mathcal{K}[s_\text{ref},z_i])^2. \label{loss}
\end{equation}
Therefore, our s-ICA method is to find out $A^{-1}$ such that $\mathcal{L}[\sum_j(A^{-1})_{ij}x_j]$ is minimized. 

There is one subtlety in s-ICA. The analytical form Eq. \ref{Kref} crucially relies on the distribution function Eq. \ref{Pref} for $s_\text{ref}$, but Eq. \ref{Pref} is correct only when the statistics is performed over a duration that is an integer times of the period $2\pi/\omega$. Thus, when we calculate $\mathcal{K}[\sum_j(A^{-1})_{ij}x_j,z]$, the statistics for all $x_j$ also needs to be carried out for the time interval being integer times of $2\pi/\omega$, otherwise $\mathcal{K}[s,z]$ always can not perfectly converge to $\mathcal{K}[s_\text{ref},z]$. Hence, it enters a paradox. Since the goal is to separate out a single frequency oscillation to determine the frequency $\omega$, $\omega$ is not known before analyzing. To solve this problem, our s-ICA method requires performing the minimization of $\mathcal{L}[s]$ iteratively. That is to say, we first choose an arbitrary period of $t$ to perform statistics for $x_j$, with which we minimize $\mathcal{L}[\sum_j(A^{-1})_{ij}x_j]$ to find out a $s^{(1)}_i(t)$. Although $s^{(1)}_i(t)$ is not a perfect single frequency oscillation because of this reason, we can still roughly determine a frequency $\omega^{(1)}_i$. Then we perform statistics for $x_j$ with time duration $2\pi/\omega^{(1)}_i$ and minimize $\mathcal{L}[\sum_j(A^{-1})_{ij}x_j]$ to find out a $s^{(2)}_i(t)$. $s^{(2)}_i(t)$ will be more close to a single frequency oscillation than $s^{(1)}_i(t)$, from which we can determine a frequency $\omega^{(2)}_i$. We can continue the procedure until we obtain a very good single frequency signal after $k$-steps and the frequency determined at each step also converges. This completes our s-ICA method. A comparison between ICA and our generalized s-ICA method is shown in Fig. \ref{Tab:comparison}.

\begin{table}[t]
\begin{tabular}
{|p{40pt}|p{85pt}|p{100pt}|}
\hline
~&ICA&s-ICA\\
\hline
Goal&Extracting source with certain feature&Extracting source that is a single frequency oscillation\\
\hline
Loss Function &Minimizing entropy&Minimizing the cumulants with respect to the cumulants of single frequency oscillation \\
\hline
Statistics&Over sufficiently long time&Over integer times of period~~(reach by iteration)\\
\hline
\end{tabular}

\caption{A comparison between ICA method and our generalized ICA method, in term of the goal, the loss function and the time period for performing statistics of the data.  }
	\label{Tab:comparison} 
\end{table}

\begin{figure}[t]
\centering
\includegraphics[width=1.0\linewidth]{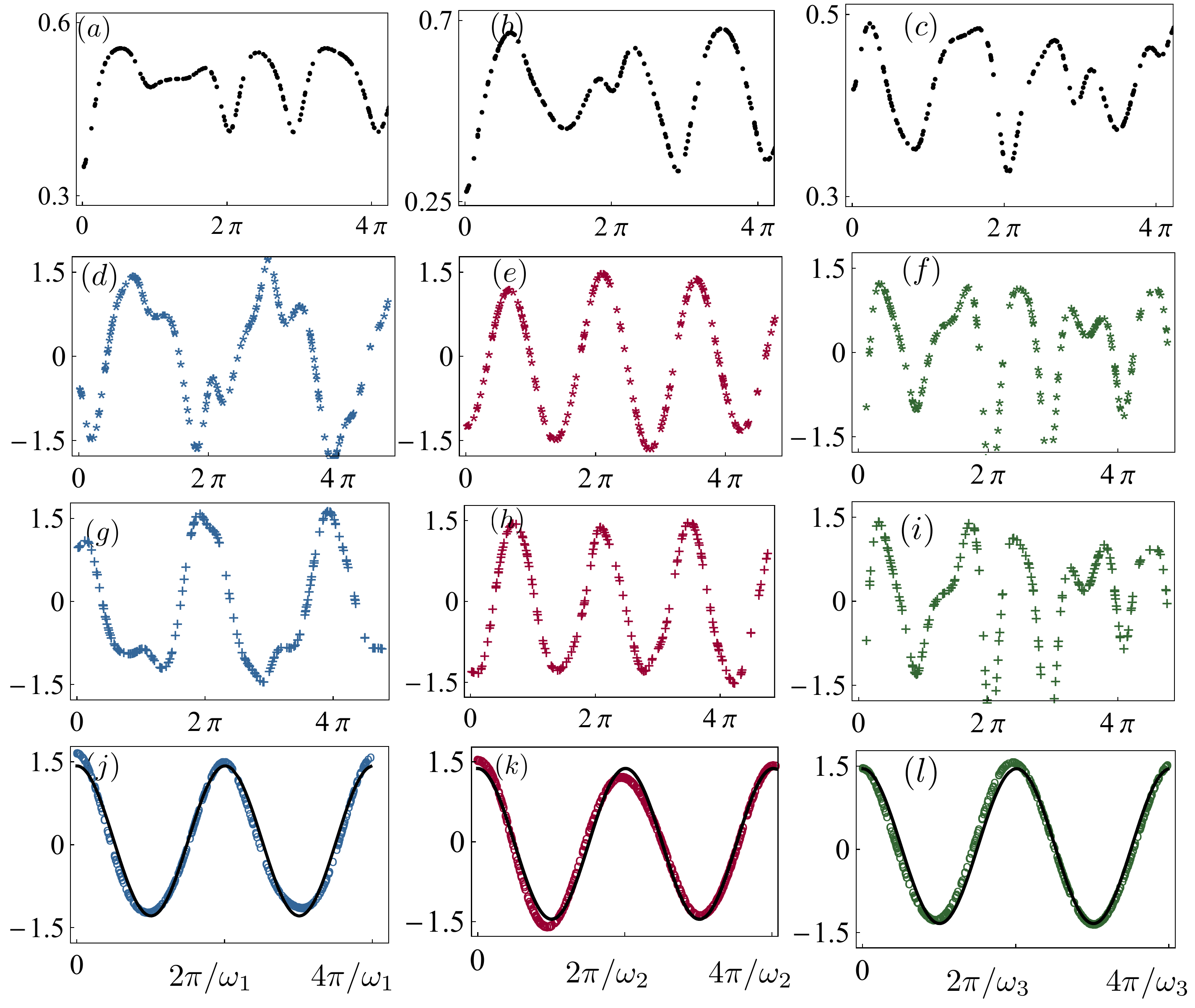}
\caption{(a-c): The original data of density oscillation of a Bose-Einstein condensate from three different detectors. Three different eigen-modes obtained by the ICA method (d-f), by our s-ICA method at the first round (g-i); and by the s-ICA method at the second round (j-l).}
\label{ICA}
\end{figure}

\textit{Analyzing Collective Modes of a BEC.} As a demonstration of our method, we apply our method to analyze the collective oscillations of a Bose-Einstein condensate (BEC) of cold atoms in a harmonic trap, which is a very common experimental measurement in cold atom experiments\cite{BECex1}. We first describe how we generate simulated data. We first consider an equilibrium Thomas-Fermi density distribution 
\begin{equation}
n_0(r)=\frac{\mu-V(r)}{g}\Theta(n_0(r)),
\end{equation} 
where $\Theta(x)=\text{max}(0,x/|x|)$ is the Heaviside function, and for simplicity,  we consider a two-dimensional geometry with a harmonic trap $V(r)=m\omega^2_\perp(x^2+y^2)/2$ and $\omega_\perp$ being the trapping frequency. $\mu$ is the chemical potential that later will be adjusted to satisfy the total number of atom conservation condition at any given time, and the total number of atom is chosen as $N=10^3$ in our simulation. $g$ is the interaction strength, and we here set $\hbar=1$, $m=1$, $\omega_\perp=1$ and $g=10$. We consider that three different modes are excited, and they are
\begin{align}
&\delta n_1=x\cos(\omega_1 t),\\
&\delta n_2=(x^2-y^2)\cos(\omega_2 t)\\
&\delta n_3=(x^2+y^2)\cos(\omega_3 t),
\end{align}
which are the dipole mode, the quadrupole mode and the breathing mode, with their frequencies being $\omega_1=1$, $\omega_2=\sqrt{2}$ and $\omega_3=2$, respectively\cite{fre1,fre2}. We also add noise that varies at different spatial location and different time, denoted by $\epsilon({\bf r},t)$. Then the simulated data of the real time density dynamics is given by
\begin{equation}
n({\bf r}, t)=\left(n_0({\bf r})+\sum\limits_{i=1}^3 f_i \delta n_i+\epsilon({\bf r}, t)\right)\Theta(n({\bf r},t)). \label{nt}
\end{equation}
where $f_i$ ($i=1,2,3$) denotes the amplitudes of the three modes. 

Here we shall also note that, although $x$, $x^2-y^2$ and $x^2+y^2$ initially are three functions orthogonal to each other, due to the boundary condition imposed by the Heaviside function and the atom number conservation condition, they are no longer orthogonal. Thus, the method like the principle component analysis\cite{PCA} also does not work very well in this case.  

\begin{figure}[t]
\centering
\includegraphics[width=0.85\linewidth]{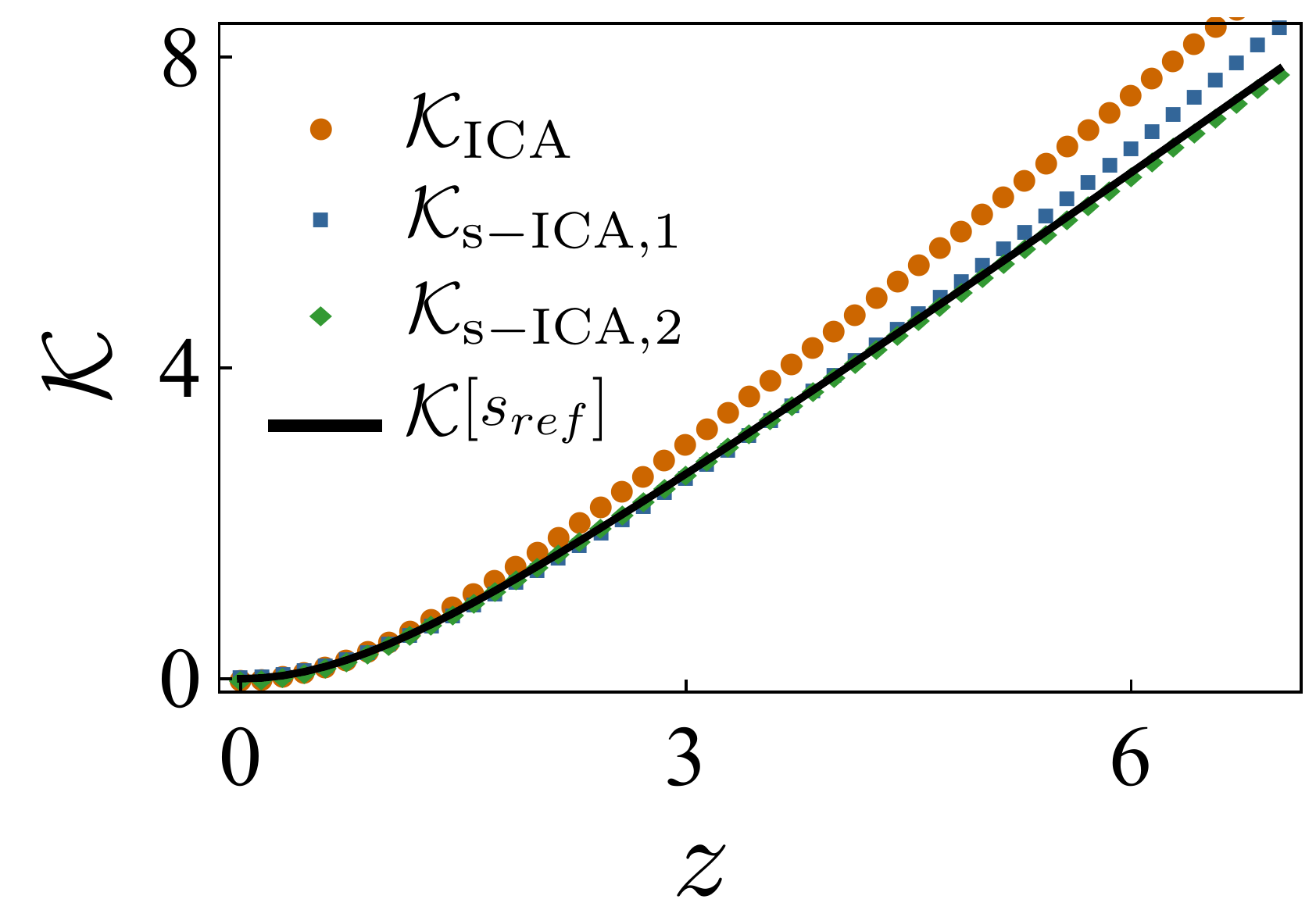}
\caption{The cumulants as a function of $z$. The solid line is for signal $s_\text{ref}$ as a perfect single frequency oscillation. The yellow circles denotes the cumulants for signal obtained by the ICA method. The blue square and the green rhombus denote the cumulants for signals obtained by our s-ICA method at the first and the second rounds, respectively. }
\label{cumulants}
\end{figure}

Now we will use the ICA method and our s-ICA method to separate out each $\delta n_i$ from data $n({\bf r},t)$ given by Eq. \ref{nt} with $f_1=0.2$, $f_2=0.2$ and $f_3=0.2$, and $\epsilon$ uniformly distributed in the range $[-0.1,0.1]$. Here we only require short-time information of couple oscillation periods, and in fact, for this application another advantage is that we only need information from few spatial points. Let us consider three locations denoted by ${\bf r}_1$, ${\bf r}_2$ and ${\bf r}_3$, and the density dynamics at these points play the role as detectors. Their density dynamics all contain these three frequencies but the coefficients are different because it depends on the spatial locations. We plot $n({\bf r_i},t)$ ($i=1,2,3$) in Fig. \ref{ICA}(a-c), which show irregular temporal behaviors. The goal of ICA or s-ICA is to find out a proper combination of them that exhibits the single frequency oscillatory behavior. 

In Fig. \ref{ICA}(d-f), we show the results of the ICA method. We can see that the quality of results are not very good. In Fig. \ref{cumulants}, we also show the cumulants calculated for the signal of Fig. \ref{ICA}(e), and one can clearly see that it derivates from $\mathcal{K}[s_\text{ref},z]$. This means that although the signal obtained by the ICA possesses certain feature, it is not yet a perfect single frequency oscillation.  

In Fig. \ref{ICA}(g-i), we show the results of the s-ICA method. The results are also not perfect, and the cumulants for the signal of Fig. \ref{ICA}(h) is also shown in Fig. \ref{cumulants} to compare with $\mathcal{K}[s_\text{ref},z]$, and one can see the the discrepancy still exists. Nevertheless, by fitting these signals, one can obtain $\omega^{(1)}_1=0.987$, $\omega^{(1)}_2=1.427 $ and $\omega^{(1)}_3=1.934$. Next, we perform statistics for two period of time duration $4\pi/\omega^{(1)}_1$ and obtain the result shown in Fig. \ref{ICA}(j), with which we obtain a new frequency $\omega^{(2)}_1=0.996$, and it only differs from the real value by $0.4\%$. For performing statistics, we have taken $400$ points in this time interval. Similarly, we preform statistics for time duration of $4\pi/\omega^{(1)}_2$, or for time duration of $4\pi/\omega^{(1)}_3$, and obtain results shown in Fig. \ref{ICA}(k) and Fig. \ref{ICA}(l), respectively, with which we determine $\omega^{(2)}_2=1.407$ and $\omega^{(2)}_3=1.989$. These results are also very close to the actually value with only $0.48\%, 0.51\%$ deviation. The cumulants for the signal of Fig. \ref{cumulants}(h) also agrees perfectly with $\mathcal{K}[s_\text{ref},z]$, as shown in Fig. \ref{cumulants}. And here we use the optimization processes to minimize the loss function Eq. \ref{loss} with Newton's method.

\begin{figure}[t]
\centering
\includegraphics[width=0.9\linewidth]{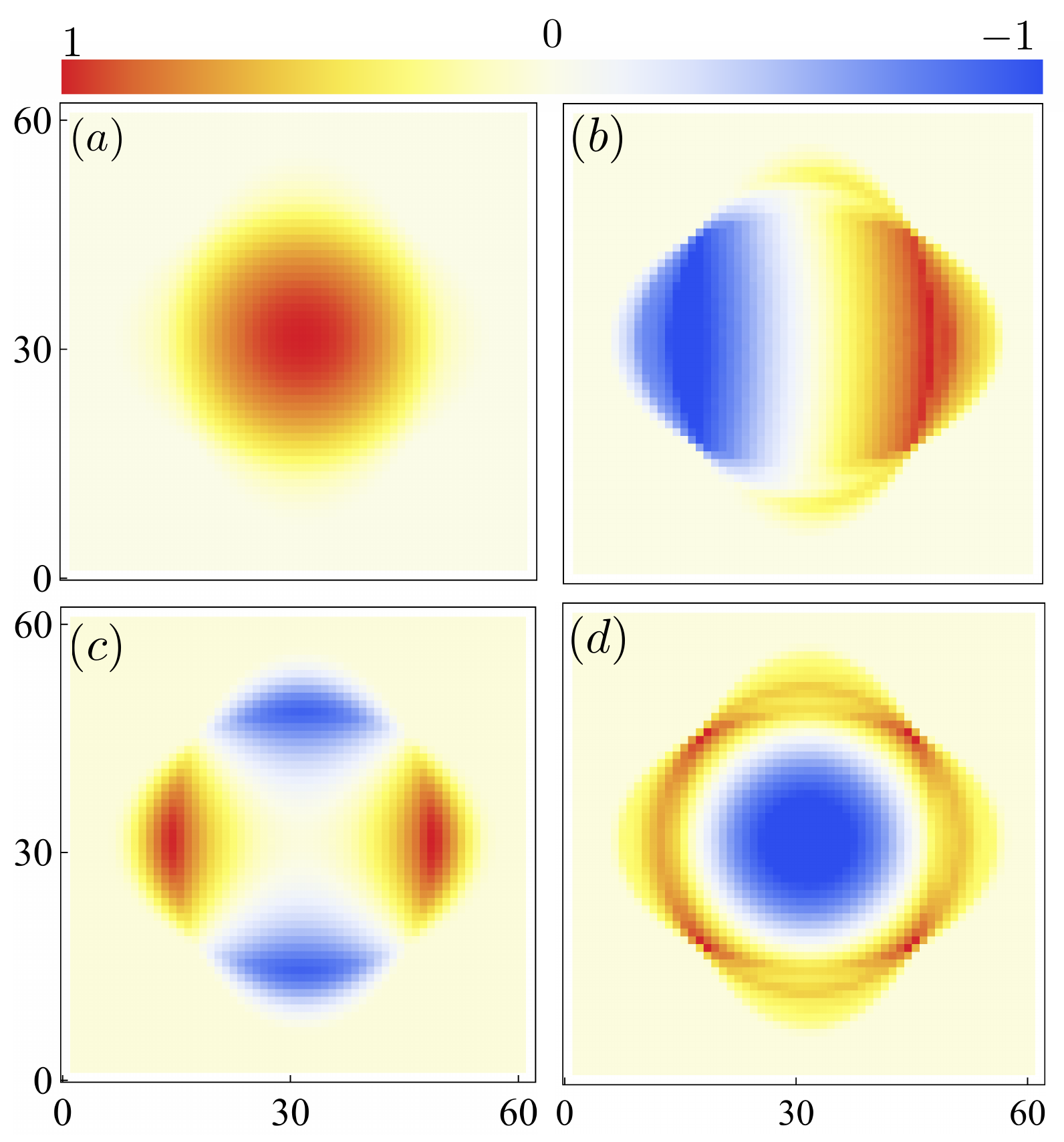}
\caption{The oscillation amplitudes at different spatial points. (a): $C_0(x,y)$; (b): $C_1(x,y)$; (c): $C_2(x,y)$; (d): $C_3(x,y)$; See text for the definition of these quantities.}
\label{coefficient}
\end{figure}

Here we should note that so far we only obtain the frequency but can determine neither the amplitude nor the damping rate. In fact, both ICA and our s-ICA have a problem that the amplitude can not be uniquely determined. In fact, signals Fig. \ref{ICA} (j-l) are all distributed between $[-\sqrt{2},\sqrt{2}]$ because the data $n({\bf r}_i, t)$ ($i=1,2,3$) have been preprocessed to be zero mean and unity variance. For instance, to obtain the amplitude, we will perform another fitting. Here we consider density dynamics from all spatial points without pretreatment, that are
\begin{equation}
n({\bf r}, t)=C_0({\bf r})+\sum\limits_{i=1}^{3}C_i({\bf r})\cos(\omega_i t).
\end{equation} 
Now all $\omega_i$ are already known from the s-ICA analysis. For each ${\bf r}$ point, there are only four fitting parameters $C_i({\bf r})$ ($i=0,\dots,3$) to fit data points of a sequence of $t$. The results of the fitting are shown in Fig. \ref{coefficient} which reveal the spatial information of each modes. In this case, Fig. \ref{coefficient}(a) is the background. Fig. \ref{coefficient}(b) shows positive amplitude in one side and negative amplitude in another side, which changes the BEC's center-of-mass and is the dipole mode. Fig. \ref{coefficient}(c) have two nodal lines and concentrates at the surface, and it is the quadruple mode. Fig. \ref{coefficient}(d) shows negative amplitudes at the center and positive amplitudes in outside, which changes the BEC's side and is the breathing mode\cite{fre2}. With similar fitting, we can also determine the damping rate of each mode.   

\textit{Outlook.} In summary, in this work we have developed a generalized ICA method to extract the eigen-energy of a quantum system from a dynamical probe. This method has advantage over other methods such as the Fourier transformation or fitting in the situation that only data of short period of time is available and is quite irregular, but our method requires accumulation of sufficient data during the time interval of few oscillation period that allows preforming accurate statistics. This is actually quite common situation in many quantum physics experiments and therefore we believe our method can find wide application in future data analysis.  

Another remark is that, although the system considered here is a quantum one, the data is the expectation value of certain observable and is a class one. It may also find its application beyond physics problems. It is of great interest to consider the quantum data and the quantum analogy of the ``cocktail party problem", and to see whether the similar ICA method can work there\cite{qICA}. 

\textit{Acknowledgment.} We thank Jia-Ming Li, Ning Sun and Ce Wang for helpful discussion. This work is supported MOST under Grant No. 2016YFA0301600 and NSFC Grant No. 11734010.

\bibliography{bibICA}

\end{document}